# Effect of the difference in ion mobilities on traveling-wave electro-osmosis

A. González[1], A. Ramos[2], P. García-Sánchez[2] and A. Castellanos[2]
1. Dpto. Física Aplicada III, Universidad de Sevilla, Camino de los Descubrimientos s/n, 41092 Sevilla (SPAIN)
2. Dpto. Electrónica y Electromagnetismo, Universidad de Sevilla, Reina Mercedes s/n, 41012 Sevilla (SPAIN)

*Abstract-* We analyze the AC electro-osmotic motion of a 1:1 aqueous solution, taking into account the difference in mobilities and diffusion coefficients between positive and negative ions. This model serves to understand the behavior of common systems as a solution of NaCl in water. We pay special attention to two cases. First, the case of slightly different mobilities, that can model a KCl solution. Second, the case of a strongly asymmetric solution, with an almost vanishing mobility, applicable to the case of a salt where the negative ion is much more massive than the positive one. For all the cases, we perform the mathematical description and linear analysis of the problem, in order to establish the dependence of the induced velocity with the frequency, wavelength and amplitude of the applied voltage.

## I. Introduction

The interaction between ac electric fields and the induced charge in the double layer formed on top of electrodes generates steady motion of aqueous solutions. This phenomenon has been termed ac electro-osmosis [1-3]. Unidirectional fluid motion is obtained when the electrolyte is placed on top of an array of micro-electrodes subjected to a traveling-wave potential [4,5]. In this work, we analyze the generated fluid motion for a 1:1 electrolyte taking into account Faradaic reactions and the difference in ion mobilities. We perform the analysis by solving the standard model of the double-layer in the linear approximation. In this way, we extend previous analysis that considered the effect of Faradaic reactions with ions of equal mobilities [6].

## II. System Description

We consider a 1:1 solution placed on top of an array of electrodes subjected to a TW signal (fig. 1). The ions have different mobilities and diffusion coefficients. One of the species (in our model, the negative one) can react on the electrodes, reversibly exchanging electrons with a neutral species, $X^- \leftrightarrow X + e^-$.

In experiments, TW electric fields are usually generated by applying a four-phase ac signal to an array of electrodes. Each electrode is subjected to an ac voltage, and there is a phase difference of 90° between consecutive electrodes. We will simplify the applied potential at the level of the electrodes by a single mode of frequency $\omega$ and wave-number $k$, i.e. we assume that the applied potential is $V_0\cos(\omega t - kx)$.

The behavior of this system can be modeled in terms of four functions: the electric potential $\phi$, and the densities of positive, negative and neutral species, $c_+$, $c_-$ and $c_0$. The electric potential obeys Poisson's equation

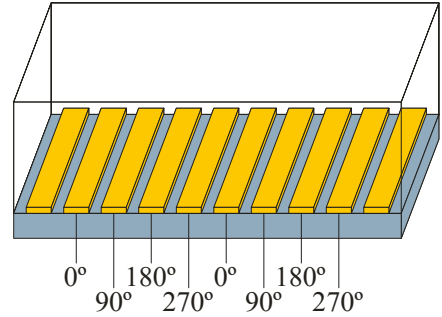

Fig 1. Model of the electrode array

$$\nabla^2 \phi = -\frac{e}{\varepsilon}(c_+ - c_-), \quad (1)$$

while the particle densities satisfy Nernst-Planck equations, neglecting convection currents

$$\frac{\partial c_+}{\partial t} = \nabla \cdot (\mu_+ c_+ \nabla \phi + D_+ \nabla c_+), \quad (2)$$

$$\frac{\partial c_-}{\partial t} = \nabla \cdot (-\mu_- c_- \nabla \phi + D_- \nabla c_-), \quad (3)$$

$$\frac{\partial c_0}{\partial t} = \nabla \cdot (D_0 \nabla c_0). \quad (4)$$

Here, the diffusion coefficients and mobilities are related through the Einstein's relation: $D_+/\mu_+ = D_-/\mu_- = k_B T/e$.

The boundary conditions at the electrodes ($y = 0$) for these functions are the continuity of the displacement vector, assuming a linear Stern layer of effective width $\lambda_s$ [6,7],

$$\frac{\varepsilon}{\lambda_s}(V_0 \cos(\omega t - kx) - \phi) = -\varepsilon \frac{\partial \phi}{\partial y}. \quad (5)$$

We assume the electrodes to be impermeable to cations.

$$\mu_+ c_+ \frac{\partial \phi}{\partial y} + D_+ \frac{\partial c_+}{\partial y} = 0, \quad (6)$$

while negative and neutral fluxes are related through the Faraday current

$$-\mu_- c_- \frac{\partial \phi}{\partial y} + D_- \frac{\partial c_-}{\partial y} = \frac{J_F}{e}, \quad (7)$$

$$D_0 \frac{\partial c_0}{\partial y} = -\frac{J_F}{e}. \quad (8)$$

This current verifies the Butler-Volmer equation [7,8]

$$\frac{J_F}{e} = K_- c_- e^{(1-\alpha)e(V_s - \phi)/k_B T} - K_0 c_0 e^{-\alpha e(V_s - \phi)/k_B T}. \quad (9)$$

Far from the electrodes ($y \to \infty$) the asymptotic behavior is given by
$$\phi \to 0, \quad c_+, c_- \to c^{eq}, \quad c_0 \to c_0^{eq}. \qquad (10)$$

The system of equations is simplified when it is expressed in terms of the charge density $\rho = e(c_+ - c_-)$ and the mean ion density $c = (c_+ + c_-)/2$. The equations can be made dimensionless using the scales $\lambda_D$ (Debye length) and $1/k$ (wavelength over $2\pi$) for the normal and tangential coordinates, $1/\omega$ for time, $k_BT/e$ for voltages, $c^{eq}$ for the charged species and $c_0^{eq}$ for the neutral species. The liquid velocity can be scaled with $\varepsilon k(k_BT/e)^2/\eta$.

Using these scales the complete system depends on a set of nondimensional parameters:
$$\gamma = \frac{D_+ - D_-}{D_+ + D_-}, \; \delta = k\lambda_D, \; \bar{\omega} = \frac{\omega\varepsilon}{\sigma(1-\gamma^2)}, \; \bar{\lambda}_s = \frac{\lambda_s}{\lambda_D},$$
$$\bar{V}_0 = \frac{eV_0}{k_BT}, \; R'_s = \frac{\sigma k k_B T}{K_- c^{eq} e^2}, \; D = \frac{D_0}{D_-}, \; N = \frac{2c^{eq}}{c_0^{eq}}. \qquad (11)$$

The parameter $\gamma$ measures the asymmetry of the solution, ranging from $-1$ (immobile cations) to $+1$ (immobile anions), for $\gamma = 0$ we have a complete symmetric electrolyte. In real situations we have that for KCl the asymmetry is small ($\gamma = -0.018$); for NaCl is moderate ($\gamma = -0.21$). If we consider that the charged species are Na$^+$ and OH$^-$, $\gamma = -0.60$, while taking H+ and Cl- the asymmetry is positive: $\gamma = +0.64$. The same happens for pure water, where $\gamma = +0.28$.

The system can be linearized expanding the quantities in powers of $V_0$. Once the system is linearized, we can assume that all first order quantities vary as traveling waves
$$A(\bar{x},\bar{y},\bar{t}) = \tilde{A}(\bar{y})e^{i(\bar{t}-\bar{x})} + \tilde{A}^*(\bar{y})e^{-i(\bar{t}-\bar{x})} \qquad (12)$$
and equations for the coefficients are obtained. The solution for each quantity is a combination of four exponentials
$$\tilde{A}(\bar{y}) = \sum_{i=1}^{4} a_i A_i e^{-\sigma_i y} \quad \operatorname{Re}(\sigma_i) > 0 \qquad (13)$$
where $a_i$ are coefficients common to all quantities and $A_i$ the components of the eigenvectors of the linear problem. The four eigenvalues are:
$$\sigma_{1,2} = \sqrt{\frac{1}{2} + i\omega + \delta^2 \pm \frac{1}{2}\sqrt{1 - 4\gamma^2\omega^2}}$$
$$\sigma_3 = \delta \qquad \sigma_4 = \sqrt{\delta^2 + \frac{i\omega(1+\gamma)}{D}} \qquad (14)$$

The corresponding eigenvectors $(\rho_i, c_i, \phi, c_{0i})$ are:
$$\mathbf{V}_1 = \left(\sigma_1^2 - i\omega - \delta^2, -i\gamma\omega, (\sigma_1^2 - i\omega - \delta^2)/(\delta^2 - \sigma_1^2), 0\right)$$
$$\mathbf{V}_2 = \left(i\gamma\omega, 1 + i\omega + \delta^2 - \sigma_2^2, i\gamma\omega/(\delta^2 - \sigma_1^2), 0\right) \qquad (15)$$
$$\mathbf{V}_3 = (0,0,1,0) \qquad \mathbf{V}_4 = (0,0,0,1)$$

The first two exponents $\sigma_1$ and $\sigma_2$ are almost equal to 1 and $(i\omega)^{1/2}$, respectively, for low frequencies and long waves. These exponents correspond to the diffuse Debye layer and the diffusion layer, respectively. When the system is symmetrical, $\gamma = 0$, there is net charge only in the diffuse Debye layer.

The effect of the asymmetry in mobilities is to connect these two scales and make the charge density extend to the diffusion layer. As a result, the computation of the electro-osmotic velocity cannot be restricted to the Debye layer and should include electrical stresses in the diffusion layer.

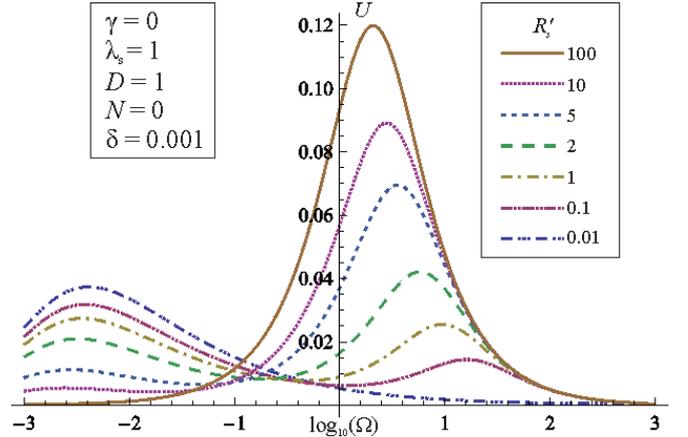

Fig 2: Electro-osmotic velocity for a symmetric solution.

The values of the coefficients $a_i$ are obtained inserting the linear combination (13) in the boundary conditions. The method is trivial, although the final results are extremely cumbersome. We have performed the algebraic manipulation using Mathematica.

Once these coefficients are calculated, we can compute the electro-osmotic velocity integrating the Stokes equations. The average tangential velocity obeys
$$0 = \frac{\partial^2 \langle \tilde{u} \rangle}{\partial y^2} - i\left(\tilde{\rho}\tilde{\phi}^* - \tilde{\rho}^*\tilde{\phi}\right) \qquad (16)$$
that, upon integration, produces the electro-osmotic velocity
$$U = i\sum_{i,j} \frac{a_i a_j^* \left(\rho_j^* \phi_i - \rho_i \phi_j^*\right)}{(\sigma_i + \sigma_j^*)^2} \qquad (17)$$

### III. RESULTS

Once we have the expression for the electro-osmotic velocity (17) we can evaluate it for different ranges of the parameters and compare the results with previous analyses and experimental data. In the following, we take $\bar{V}_0 = 1$.

For the case of a symmetric solution ($\gamma = 0$) we can examine the change in the velocity when the reaction resistance $R'_s$ varies from infinity (no Faraday currents) to 0 (*facile kinetics* regime). We plot the velocity as a function of the scaled frequency $\Omega = \omega(1-\gamma^2)/\delta = \varepsilon\omega/\sigma k\lambda_D$ (Fig. 2). Typical ac electroosmotic flows are obtained for $\Omega \sim 1$ [1-3]. For high resistance, there is a single maximum close to $\Omega = 1 + \lambda_s$ (the ideally polarisable limit). As the magnitude of the Faraday currents grow, this maximum diminishes and a new maximum appears at lower frequencies, which is related to the Warburg impedance of the Randles circuit [6,8].

For a strongly asymmetrical solution where positive ions have a much higher diffusivity than the negative, reacting ones, the behavior changes (Fig. 3). The second maximum, associated to the Warburg impedance, moves to higher frequencies, while at extremely low frequencies there appears a reversal flow, where the electro-osmotic velocity is in opposite direction to the traveling wave.

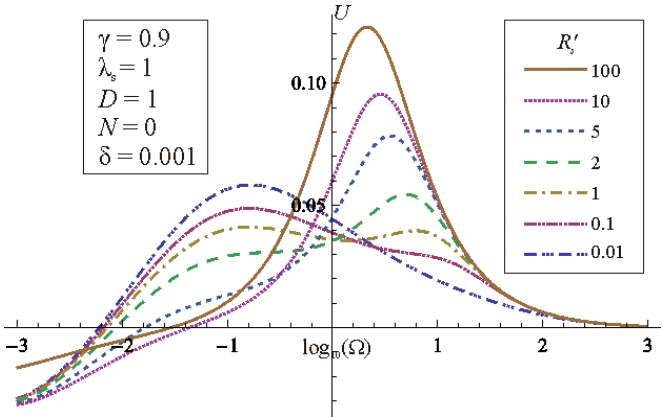

Fig 3: Electro-osmotic velocity for a asymmetric solution, where negative ions have a small diffusivity.

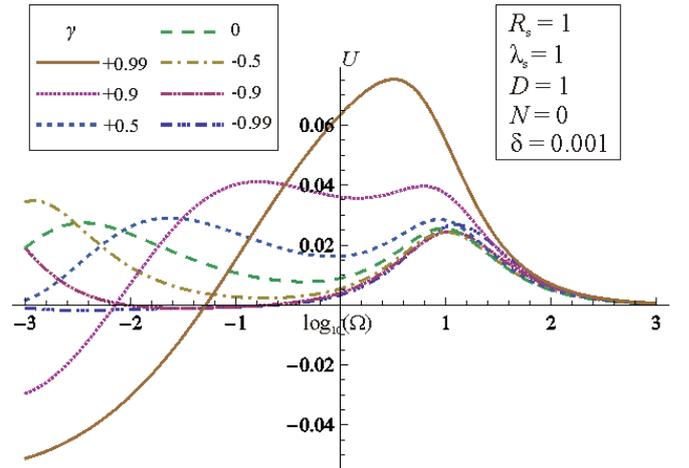

Fig 5. Electro-osmotic velocity for different values of the asymmetry

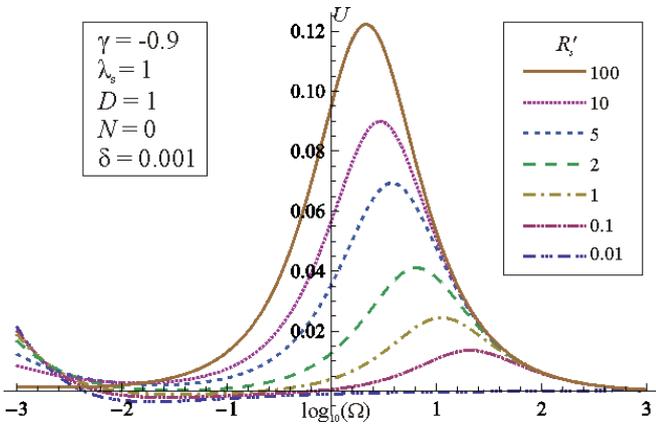

Fig 4. Electro-osmotic velocity for a asymmetric solution, where negative ions have a high diffusivity.

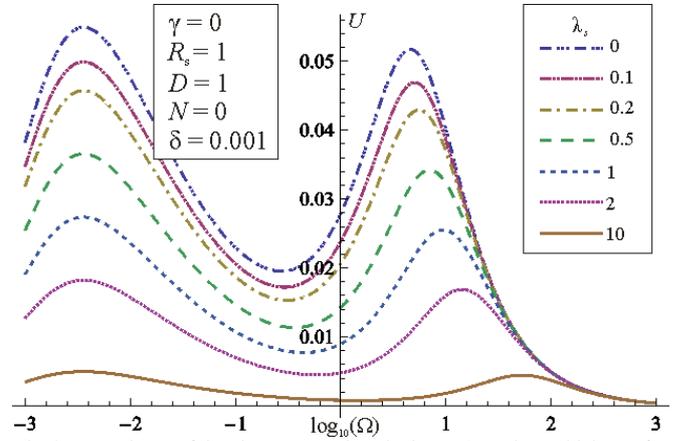

Fig 6. Dependence of the electro-osmotic velocity on Stern layer thickness for a symmetric solution.

If the negative ions have much higher diffusivity than cations, i.e. γ is close to −1, (Fig. 4) the secondary maximum moves to even lower frequencies, disappearing in practice (as the dimensional frequencies would be below 1Hz). In the facile kinetics regime, the electro-osmotic velocity disappears, except for a very small reverse flow.

We can show the dependency on γ by showing the electro-osmotic velocity for a given value of $R'_s$ and different values of the asymmetry (Fig 5).

The previous figures have a constant Stern layer thickness (taken as equal to the diffuse layer thickness). When we consider the effect of changing this thickness, we obtain that for a symmetrical electrolyte the main effect is a reduction on the electro-osmotic velocity (Fig. 6). However, when the solution is asymmetric and the negative ions have a very high diffusivity (Fig. 7) the velocity can change its direction, being this reverse flow of small amplitude for high values of the Stern layer thickness. This effect is relatively more important when the Stern layer is thick compared with the Debye layer.

If the negative ions have a small diffusion coefficient (Fig. 8) the effect is much smaller.

### IV. CONCLUSIONS

We have performed a linear analysis of traveling wave electro-osmosis for an aqueous solution where the charged species can have very different mobilities and diffusion coefficients, and Faraday currents are possible. The resulting expression for the electro-osmotic velocity, $U$, depends on a large number of parameters. We have considered the dependence on the frequency of this velocity, for different values of the parameters.

Our results show that, in general, a combination of Faradaic currents and asymmetric mobilities reduces the strength of the electro-osmotic velocity and, in some cases, leads to reverse flow. The reverse flow observed in some experiments may be related to these two factors [4,5].

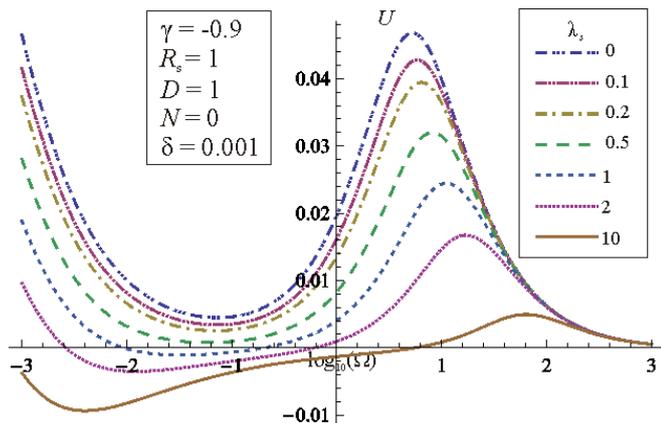

Fig 7. Dependence of the electro-osmotic velocity on Stern layer thickness for an asymmetric solution where negative ions a very high diffusivity.

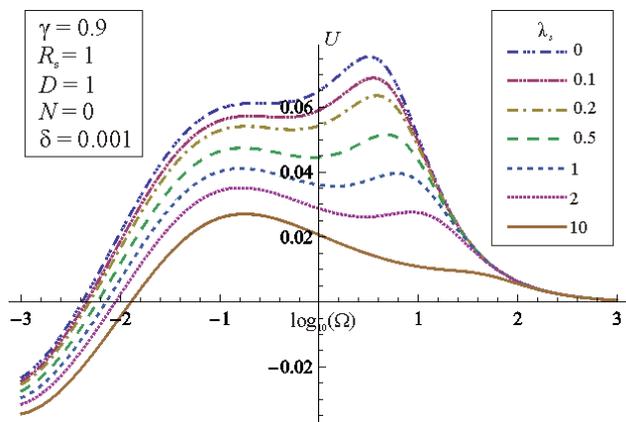

Fig 8. Dependence of the electro-osmotic velocity on Stern layer thickness for an asymmetric solution where negative ions a very small diffusivity.

ACKNOWLEDGEMENTS

This work has been supported by the Spanish government agency DGCyT and Junta de Andalucía under contracts FIS2006-03645 and FQM-241, respectively.

REFERENCES

[1] Ramos A, Morgan H, Green NG, Castellanos A, J Colloid Interf Sci, **217**, 420 (1999).
[2] González A, Ramos A, Green NG, Castellanos A, Morgan H, Phys Rev E **61**, 4019 (2000)
[3] Green NG, Ramos A, González A, Morgan H, Castellanos A, Phys Rev E **61**, 4011 (2000)
[4] Ramos A, Morgan H, Green NG, González A, Castellanos A, J Appl Phys **97**, 084906 (2005)
[5] García-Sánchez P, Ramos A, Green NG, Morgan H, IEEE Trans Dielectr Electr Insulat **13**, 670–677 (2006)
[6] Ramos A, González A, Castellanos A, García-Sánchez P, J Colloid Interf Sci, **309**, 323–331 (2007)
[7] Bonnefont A, Argoul F, Bazant MZ, J Electroanal Chem, **500**, 5261 (2001)
[8] Bard AJ, Faulkner LR, Electrochemical Methods: Fundamentals and Applications, second ed., Wiley, New York, 2001.